\documentclass[a4paper,aps,prb,twocolumn,preprintnumbers,amsmath,amssymb,superscriptaddress]{revtex4-1}
\usepackage{amsmath}
\usepackage{verbatim}
\usepackage{graphicx,epstopdf}
\usepackage{hyperref}

\usepackage{color}

 \newcommand{\ket}[1]{\left|#1\right\rangle} 
 \newcommand{\bra}[1]{\left\langle#1\right|} 

\usepackage{xspace}
\usepackage{xparse}
\usepackage{amsfonts,amssymb,amsthm,amscd} 
\def\subwavelength{sub-diffraction\xspace}

\def\upSpin{{\uparrow}}
\def\downSpin{{\downarrow}}
\def\up{{\Uparrow}}
\def\down{{\Downarrow}}
\def\OmegaP{\Omega_p\xspace}
\def\OmegaC{\Omega_c\xspace}
\def\deltaDown{\delta_\down}
\newcommand{\rs}{\rm \scriptscriptstyle}

\begin{document}
 
\title{Optical control over bulk excitations in fractional quantum Hall systems}

\author{Tobias Gra{\ss}}
\affiliation{Joint Quantum Institute, NIST and University of Maryland, College Park, Maryland, 20742, USA}
\author{Michael Gullans}
\affiliation{Department of Physics, Princeton University, Princeton, New Jersey, 08544, USA}
\author{Przemyslaw Bienias}
\affiliation{Joint Quantum Institute, NIST and University of Maryland, College Park, Maryland, 20742, USA}
\author{Guanyu Zhu}
\affiliation{Joint Quantum Institute, NIST and University of Maryland, College Park, Maryland, 20742, USA}
\author{Areg Ghazaryan}
\affiliation{Department of Physics, City College, City University of New York, New York, NY 10031, USA}
\author{Pouyan Ghaemi}
\affiliation{Department of Physics, City College, City University of New York, New York, NY 10031, USA}
\affiliation{Department of Physics, Graduate Center, City University of New York, New York, NY 10031, USA}
\author{Mohammad Hafezi}
\affiliation{Joint Quantum Institute, NIST and University of Maryland, College Park, Maryland, 20742, USA}
\affiliation{Department of Electrical Engineering and Institute for Research in Electronics and Applied Physics, University of Maryland, College Park, MD 20742, USA}

\begin{abstract}
Local excitations in fractional quantum Hall systems are amongst the most intriguing objects in condensed matter, as they behave like particles of fractional charge and fractional statistics. In order to experimentally reveal these exotic properties and further to use such excitations for quantum computations, microscopic control over the excitations is necessary. Here we discuss different optical strategies to achieve such control. First, we propose that the application of a light field with non-zero orbital angular momentum can pump orbital angular momenta to electrons in a quantum Hall droplet. In analogy to Laughlin's argument, we show that this field can generate a quasihole or a quasielectron in such systems.  Second, we consider an optical potential that can  trap a quasihole, by repelling electrons from the region of the light beam.  We simulate a moving optical field, which is able to control the position of the quasihole. This allows for imprinting the characteristic Berry phase which reflects the fractional charge of the quasihole.
\end{abstract}

\maketitle
\section{Introduction}
Since the discovery of topological matter in two dimensions \cite{tsui82,laughlin83} it has become clear that the distinction between fermions, as particles which obey the Pauli exclusion principle, and bosons, as particles which do not, is incomplete on the level of emergent particles. Instead, topological many-body systems can host also quasiparticles, so called anyons with intermediate quantum-statistical behavior.  \cite{leinaas77,wilczek82}. In many respects, an anyon behaves like the fraction of a particle, and accordingly it possesses fractional quantum numbers. For instance, electronic systems in the fractional quantum Hall regime host quasiparticles whose electric charge is only a fraction of the electron's charge. If two identical anyons are exchanged, their wave function may acquire a $U(1)$ phase, which in contrast to the case of bosons and fermions is not restricted to integer multiples of $\pi$. An even more exotic type of anyons are the non-Abelian ones \cite{moore-read}: they have a characteristic number of (quasi-)degenerate ground states, and under particle exchange a state in this manifold can evolve into another one. Importantly, such mixing is not possible under local perturbations, which has triggered the hope for an exciting technological application, namely a robust quantum memory. The quantum information stored in the topologically protected state of the anyons can be processed by the braiding of non-Abelian anyons, possibly allowing for fault-tolerant quantum computing \cite{nayak08}. The first step to achieve this goal is to gain control over quasiparticles in fractional quantum Hall systems.

The standard way of creating fractional excitations is by tuning the magnetic field strength and/or the electrostatic backgate potential. Current schemes for detecting anyonic behavior are based on transport measurements in interferometric devices \cite{camino05,mcclure12,willett13}. However, there are also different optical techniques which can be used to probe quantum Hall physics beyond electronic transport measurements: Since the early days of quantum Hall physics, the light emission from quantum Hall samples has been measured \cite{goldberg90,turberfield90} in order to probe the interaction between electrons and holes \cite{macdonald90,cooper97,wojs00}. In addition to emission spectra, also the elastic and inelastic scattering of light has been detected \cite{levy16}. Recently, a novel spectroscopic approach with improved energy resolution has been achieved by bringing a GaAs quantum well into a cavity, and detecting polariton resonances via light reflection \cite{ravets18}. Landau level transitions in graphene have been probed by infrared absorption spectroscopy \cite{jiang07,orlita08}, and Raman spectroscopy \cite{goler12,faugeras15}. Photocurrent measurements in graphene have combined optical probing with transport measurements \cite{nazin10,olivier}. Moreover, using a scanning single-electron transistor \cite{Ilani2004} or a scanning tunneling microscope \cite{morgenstern03,matsui2005,hashimoto08,luican-mayer14}, the local density of states has been detected for graphene in the quantum Hall regime. The resolution of these measurements allows for identifying single quasiparticles, and this technique has recently been suggested for the direct imaging also of fractional quasiparticles \cite{papic18}.

At the same time, there have also been remarkable advances in optical control and manipulation of synthetic many-body systems. For atomic quantum Hall droplets, it has been proposed to create anyons via AC Stark shift, and to directly observe their dynamical behavior \cite{paredes01,julia-diaz12,grass12}. In such systems, spectroscopic properties can also reveal the fractional statistics of excitations \cite{cooper15}. In optical lattices, adiabatic flux insertion is suited to grow fractional quantum Hall states \cite{grusdt14}, or to create anyonic excitations \cite{wang18}. Angular-momentum resolved spectroscopy of emitted light has been suggested as a tool to gain microscopic insight to a photonic quantum Hall system \cite{umucal17}. Exploiting light beams with orbital angular momentum has been proposed for the engineering of polaritonic fractional quantum Hall systems \cite{ivanov18}.

\begin{figure}[t]
\centering
\includegraphics[width=0.45\textwidth]{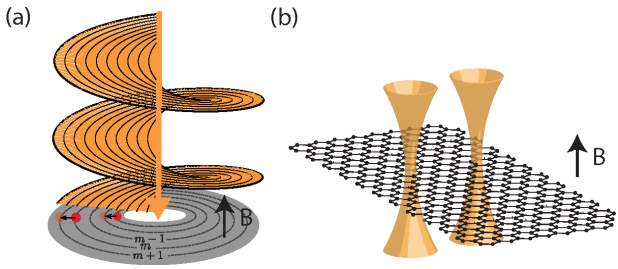}
\caption{\label{scheme} 
{\bf Two schemes to optically prepare quasi-particles in a FQH system.} (a) Synthetic flux insertion: Light with orbital angular momentum couples to a quantum Hall system on a Corbino disk, and shifts (quasi-)particles through the annulus. Since only entire electrons may flow through a wire connecting inner and outer edge,  a fractional quantum Hall system at filling $\nu=1/q$ requires $q$ pumping cycles for a measurable signal. (b)  Light-induced potentials: Local light beams create an optical Stark shift which is suited to trap quasiparticles in a FQH system.
}
\end{figure}

In the present paper, we apply a quantum optics toolbox to manipulate electrons in quantum Hall liquids.  A major advantage of optical methods is their versatility. For instance, while manipulating an electronic material with a gate potential requires built-in contacts, optical potentials could have less hardware requirements. Moreover, compared to transport measurements, optical schemes can be suited for local probes, and the position of an optical potential can flexibly be tuned. These properties suggest that optical techniques may become particularly useful for braiding quasiparticles. In this paper, we present two different schemes to create and manipulate quasiparticles in electronic fractional quantum Hall systems, as schematically shown in Fig.~\ref{scheme}: (a) Synthetic flux insertion creates quasiparticles, and (b) light-induced potentials are able to trap quasiparticles.

\textbf{(a) Synthetic flux insertion:} In this scheme, presented in Sec. \ref{pump}, we exploit the orbital angular momentum of light to synthesize the insertion of a magnetic flux, and to create individual quasiholes or quasielectrons. Specifically,  we use a pulsed light field with a non-zero orbital angular momentum, and coherent light-matter interactions to pump the electrons into a state with the angular momentum shifted by the value of the photons' orbital angular momentum. From the conceptual point of view, this process is equivalent to adding/removing a magnetic flux into/from the system. Therefore, each light pulse can be designed to exactly produce one quasihole or one quasielectron, if the orbital angular momentum of the light field is $\pm1$.

Details of the optical coupling depend on the material:
(i) For Dirac materials like graphene, we consider a single optical transition from the Landau level at the Fermi surface to an empty Landau level. Such a selective coupling is enabled by the anharmonicity of the relativistic energy spectrum, in contrast to systems with quadratic dispersion, e.g., GaAs.  If the electron and the photon exchange orbital angular momentum, such a transition can be used to change the angular momentum of an electron. By timing the pulse duration, such that it matches the value $\pi$ (in units given by the inverse Rabi frequency), we can coherently increase (decrease) the angular momentum of all electrons by one, and thereby, produce a quasihole (quasielectron). 
(ii) In systems with quadratic dispersion, we consider a Raman-type coupling between two spin manifolds in the conduction band, and the valence band.
This approach requires spin-orbit coupling in the valence band, as found in GaAs. Exploiting a STIRAP-like protocol, cf. Ref. \onlinecite{vitanov17}, the Raman beams coherently flip the spin of the (spin-polarized) fractional quantum Hall state without producing excitations from the valence band. As in the case of graphene, it is again possible to increase (decrease) the angular momentum of each electron by using light with orbital angular momentum. Both protocols are robust against disorder provided the  timescale for the optical transfer process is fast compared to the disorder potential.

Our optical method is particularly unique as it provides an experimentally applicable and well controlled method to generate quasielectrons in a similar setting as the quasiholes. Despite their apparent similarity, there is a fundamental distinction between the two types of quasiparticles, and this can be understood from the fact that the quasihole state is an eigenstate of the parent Hamiltonian for the Laughlin state in the presence of an additional repulsive potential. In contrast, adding an attractive potential in place of the repulsive one does not generally produce the quasielectron state, due to the fact that the position of the potential might be already occupied by another electron \cite{Hansson2009,Nielsen2018}. 

\textbf{(b)  Light-induced potentials:} In this scheme, we propose that by using an off-resonance light field, one can generate an AC Stark shift to produce a local potential, as presented in Sec. \ref{pin}. By choosing the frequency of the light field to be larger than the one of the corresponding Landau level transition, we show that the resulting repulsive potential can stabilize the system with a single quasihole. The optimal size of the trap corresponds to a situation where  the potential width is of the order of the magnetic length. However, within the quantum Hall regime, this length scale is usually smaller than  the wavelength of light, which sets the minimal trap size. Improvements on the potential can be achieved by an alternative sub-wavelength trapping scheme. Specifically, by coupling three electronic levels, e.g., from two spin levels and the valence band, it is possible to engineer optical potentials below the diffraction limit. Furthermore, we  show that by adiabatically moving the optical potential, the electronic wave function acquires a Berry phase which reflects the fractional charge of the excitation. By explicitly simulating the dynamics of a small system, we determine the maximum speed at which the potential can be moved without resulting in non-adiabatic behavior. Our simulation demonstrates that optical traps may become a useful tool for the braiding of anyons.

Although our optical schemes can be applied to different fractional or integer quantum Hall phases, the present paper focuses on systems in the Laughlin phase \cite{laughlin83}. Despite its relatively simple form, the Laughlin wave function supports fractional quasiparticles, and it captures very well the ground state of electrons at filling 1/3. In the remainder of this introductory section, we briefly discuss some general properties of the Laughlin wave function and of its excitations, which will be used in this work. 

The Laughlin wave function is given by 
\begin{align}
\Psi_{\rm L}\sim\prod_{i,j} (z_i-z_j)^3 \exp[-\sum_i |z_i|^2/(4 l_{\rm B})],
\end{align}
where $z_j=x_j- i y_j$ are the spatial coordinates of the $j$th electron, with $l_{\rm B} = \sqrt{\hbar/(eB)}$ being the magnetic length in a field of strength $B$. The wave function assumes that electrons are confined to the lowest Landau level, denoted by a Landau level index $n=0$. We note that within the $n=0$ Landau level, there is no difference between graphene and semiconducting materials, except for an additional valley degree of freedom in graphene. In the Laughlin state, spin and valley degrees of freedom are assumed to be fully polarized.

For $N$ electrons, the $z$-component of total angular momentum in the Laughlin state is $L_N=\frac{3}{2}N(N-1)$. In the thermodynamic limit, this quantum number is replaced by the filling factor $\nu$, that is, by the ratio between the number of electrons, and the number of states within a Landau level. The filling factor is also well-defined in compact geometries such as the torus. The Laughlin wave function corresponds to a filling $\nu=1/3$. 

One distinguishes between low-energy excitations at the edge and in the bulk of a Laughlin system. Excitations at the edge are gapless deformations, which increase the angular momentum slightly (by a value $\ll N \hbar$). The anyonic quasiparticles which we are interested in here are excitations in the bulk.
They are gapped excitations which appear as fractional electrons (``quasielectrons'') or fractional holes (``quasiholes''), that is, as a local increase or decrease of the charge density. The wave function of a quasihole at position $\xi$ is obtained by multiplying the Laughlin wave function with a prefactor $f_{\rm qh}^\xi=\prod_{i=1}^N (z_i-\xi)$. From this expression it is seen that a quasihole increases the $z$-component of total angular momentum by $\mathcal{O}(N)$ above the Laughlin value (in units $\hbar$). In contrast, for a quasielectron in the lowest Landau level located at a position $\xi$ the Laughlin wave function should be multiplied by $f_{\rm qe}^\xi=\prod_{i=1}^N (\partial_{z_i}-\xi)$, where the derivative does not act on the exponential factor of the Laughlin wave function. Obviously, the quasielectron has the opposite effect on total angular momentum compared to quasihole.

Within the lowest Landau level, the coordinate $z_i$ can be replaced by the operator $b_i^\dagger$ which raises the angular momentum of an electron. With this, we can re-write the quasihole state as 
\begin{align}
\label{bqh}
\Psi_{\rm qh}^{\xi} \sim  \prod_{i=1}^N \left(b_i^\dagger-\xi\right) \Psi_{\rm L},
\end{align}
Choosing the quasihole position to be in the center, $\xi=0$, this expression shows that the quasihole state can be produced by shifting each electron into the next angular momentum orbital. A similar expression can be obtained for the quasielectron by replacing $b_i^\dagger$ with $b_i$, which shows that in this case the angular momentum of each electron should be decreased by one. This observation outlines the strategy which we will pursue in the following section in order to generate Laughlin quasiparticles.

\section{Synthetic flux insertion}
\label{pump}

\begin{figure}[t]
\centering
\includegraphics[width=0.48\textwidth, angle=0]{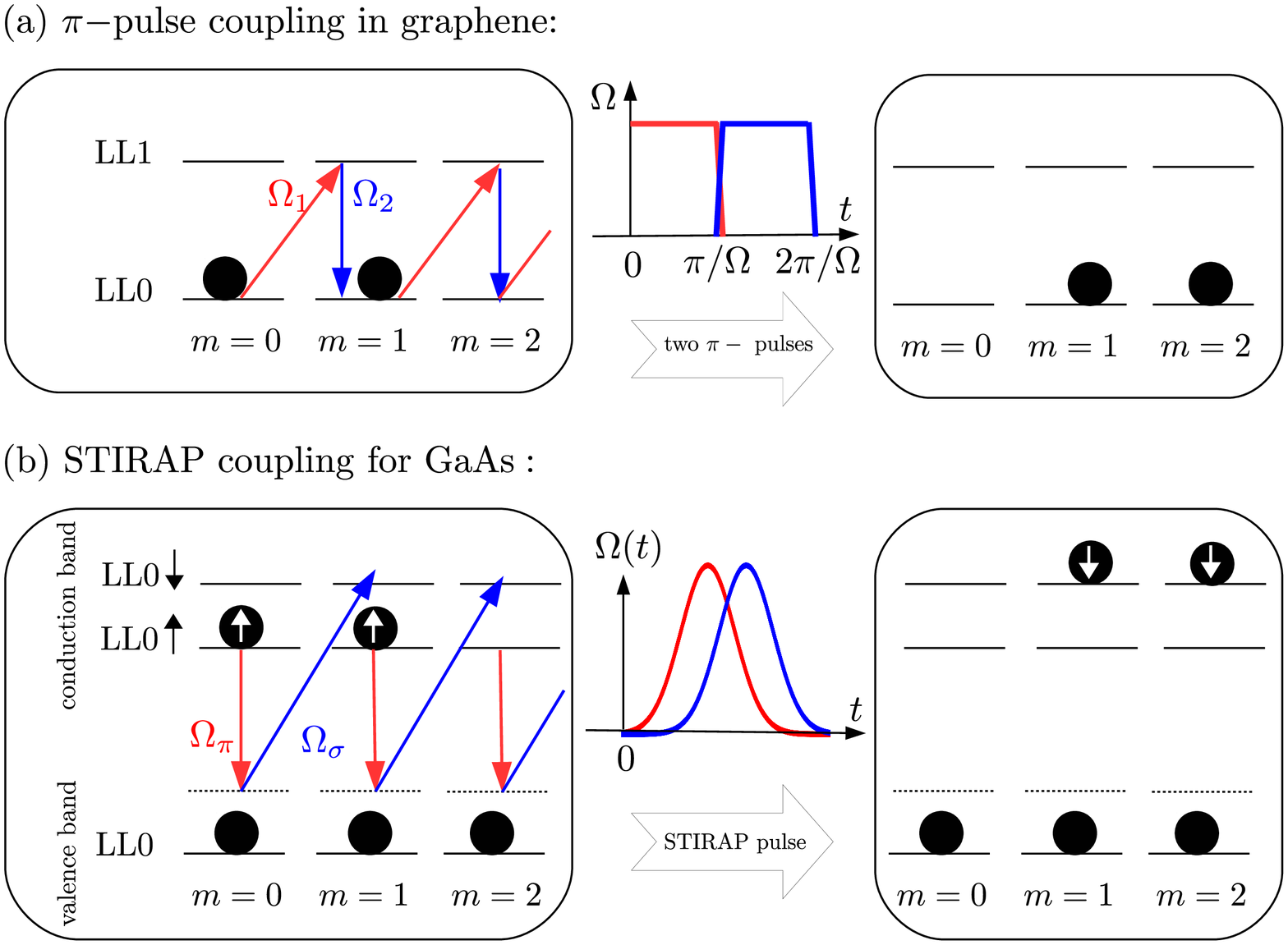}
\caption{\label{fig:raman} {\bf Different schemes for generating quasiholes.} (a) Coupling scheme for graphene: Electrons from the Fermi surface at LL0 are shifted to an empty Landau level (LL1) by a $\pi$-pulse. If the beam carries orbital angular momentum, it will also act on the orbital quantum number, $m\rightarrow m+1$. A second $\pi$-pulse is applied to remove the Landau level excitation, while leaving the angular momentum increased. The final state is a quasihole excitation at the Fermi level. (b) Coupling scheme for GaAs quantum well: A Raman-like coupling, consisting of a $\pi$- and a $\sigma$- polarized light beam, couple two spin Landau levels in the conduction band to the valence band. Using the shown STIRAP-like timing of the pulses, we can flip the spin of all conduction band electrons, while avoiding excitations from the valence band. If one of the light beams carries orbital angular momentum, the spin flip is combined with an orbital shift, $m\rightarrow m+1$. The final state is a quasihole excitation within the spin-excited Landau level.}
\end{figure}

In this section, we present two approaches to synthesize the insertion of a flux, i.e. to add quantized angular momenta to the electronic system. In both approaches, we achieve this by applying a light field with orbital angular momentum. Our first approach, presented in Sec. \ref{graphene}, uses several $\pi$-pulses, which resonantly couple two Landau levels. This approach is best suited for Dirac materials such as graphene with an anharmonic Landau level spectrum. The schematics of our approach is illustrated in Fig. \ref{fig:raman}(a). The proposed coupling brings an electron from the Fermi surface into an empty Landau level, so the action of the coupling is to change the  Landau level index by one:  $n \rightarrow n+1$. Simultaneously, the coupling increases the electron angular momentum by $\hbar$, i.e. the orbital quantum number within the Landau level increases by one: $m \rightarrow m+1$. By a proper timing of this coupling (that is, by applying a $\pi$-pulse), all electrons can coherently be transferred, resulting in a quasihole state according to Eq. (\ref{bqh}) within a higher Landau level. To remove the Landau level excitation, one can apply a second $\pi$-pulse at constant angular momentum.  In this Section, we perform a numerical simulation of the system dynamics which shows that decoherence due to electron-electron interactions is small, if the Rabi frequency is on the order of the Coulomb interactions ($\sim1$~eV). Spontaneous emission from the excited Landau level can then be neglected, as lifetimes on the order of picoseconds are much longer than the duration of a $\pi$-pulse. 

The second approach, presented in Sec. \ref{GaAs}, is based on a three-level scheme, as shown in Fig. \ref{fig:raman}(b). Here, a Raman-type coupling between two spin Landau levels in the conduction band leads to a spin flip. An important ingredient to enable the Raman transition is spin-orbit coupling in the valence band, which can be found in prominent quantum Hall materials, including GaAs. As in our first approach, angular momentum transfer from the photons to the electrons generates the desired orbital shift, $m \rightarrow m+1$. This approach produces a quasihole state within the spin-reversed Landau level. Since interactions are spin-independent, this scheme is free from decoherence due to interactions. Moreover, lifetimes of spin excitations are extremely long (on the order of nanoseconds) \cite{kikkawa98,ohno99}, so the final state is sufficiently stable.  Excitation of valence-band electrons can be avoided by applying a detuned STIRAP protocol. 

In Sec. \ref{sec:corbino}, we discuss an experimental proposal to measure the fractional charge of an anyon. The main idea is that by increasing the angular momentum of each electron by $\hbar$, a charge $e/q$ is pumped through the system, with $q=1/\nu$, which is set by the filling factor $\nu$.

\subsection{ $\pi$-pulse coupling in graphene \label{graphene}}
We consider an optical coupling between the fractionally filled Landau level $n$ at the Fermi surface to an empty Landau level $n'$. Such a selective coupling is possible in Dirac materials such as graphene, as they exhibit a non-equidistant Landau level spectrum. The selection rules for circularly polarized light, $|n| \leftrightarrow \pm(|n|\pm 1)$, determine optically allowed transitions, cf. Ref. \onlinecite{jiang07}.
For concreteness, we will focus on graphene at a fractionally filled $n=0$ level, and consider a coupling to $n'=1$. We note that both Landau levels support Laughlin-like ground states \cite{areg,toke06,amet15} at filling $\nu=1/3$. If the light also carries orbital angular momentum $\ell$, the selection rule regarding the orbital quantum number $m$ is given by $|\Delta m|=\ell$, see Ref. \onlinecite{gullans17}. 

In the rotating frame, such coupling is described by a time-independent Hamiltonian \cite{areg}:
\begin{align}
\label{H0}
 H_0 = \sum_{m} \left[ \hbar \delta c_{m,1}^\dagger c_{m,1}  + \hbar \frac{\Omega_m}{2} \left( c_{m+\ell,1}^\dagger c_{m,0}  + {\rm h.c.} \right) \right].
\end{align}
The operators $c_{m,n}^\dagger$ ($c_{m,n}$) create (annihilate) electrons in the $m$th orbital of the $n$th Landau level, $\delta$ is the detuning of the laser field from the Landau level transition frequency, and $\Omega_m= \int {\rm d}^2r \ \bra{m+\ell,1} {\bf E}({\bf r}) \cdot {\bf r} \ket{m,0}$ is the Rabi frequency of the optical transition $m \rightarrow m+\ell$ in the (3D) electric field ${\bf E}({\bf r})$ within the 2D-plane ${\bf r}$. 

In the following, we will take $\Omega_m=\Omega$, that is, a constant for all orbitals $m$. This assumption is not strictly valid for any system size, since in order to carry orbital angular momentum, the light beam must have a vortex line somewhere, and orbitals which are localized near the vortex line will experience a weaker Rabi frequency than others. In a large enough system only few orbitals are affected from the vortex, and our assumption of a $m$-independent Rabi frequency holds approximately. The assumption becomes more rigorous, if one considers a Corbino disk geometry, i.e. a disk with a hole in the center, such that the hole may coincide with the vortex of the light beam.

Considering the time evolution under $H_0$ in the weakly detuned limit, $\delta\rightarrow 0$, the electrons are found to perform Rabi oscillations between orbitals $\ket{n=0,m}$ and $\ket{n=1,m+\ell}$ with period $T=2\pi/\Omega$. That is, if initially all electrons were in the $n=0$ level, they will be flipped into $n=1$ after a time $t=T/2$.
A light field which is properly timed, i.e. a $\pi$-pulse, will therefore modify the initial $N$-electron state, $\ket{\Psi(0)}$, in the following way:
\begin{align}
\label{qhprime}
\ket{\Psi(T/2)} &= e^{-i \pi H_0/\Omega }\ket{\Psi(0)} = \prod_{i=1}^N [\tilde a_i^\dagger (\tilde b_i^\dagger)^\ell] \ket{\Psi(0)} \nonumber \\ &
\equiv \ket{\Psi_{\rm qh}'}.
\end{align}
Here, $\tilde a_i^\dagger$ and $\tilde b_i^\dagger$ denote the operators which raise the Landau level index $n$ and the orbital quantum number $m$:
\begin{align}
\label{btilde}
 \tilde a^\dagger \equiv \sum_{n,m} \ket{n+1,m}\bra{n,m},  \ \ \  \tilde b^\dagger \equiv \sum_{n,m} \ket{n,m+1}\bra{n,m}.
\end{align}
For $\ell=1$, the state defined in Eq.~(\ref{qhprime}) describes a quasihole excitation similar to the one defined in Eq.~(\ref{bqh}). However, Eq.~(\ref{bqh}) defines the quasihole by applying ladder operators to the ground state wave function, whereas Eq.~(\ref{qhprime}) uses projection operators $\tilde b^\dagger$. The ladder operators differ from the projection operators by a normalization factor $\sqrt{m+1}$. The effect of these normalization factors in the ladder operators is minor for small systems and vanishes in the thermodynamic limit, as we show in the appendix. Thus, the orbital shift in Eq.~(\ref{qhprime}) produces a quasihole excitation. In addition to the orbital shift, the $\pi$-pulse also increases the Landau level index of each electron. Such a projection of the Laughlin state and its excitations into higher Landau levels is a straightforward generalization which has been discussed in Refs. \onlinecite{macdonald84,macdonald86}. 

Having established that an idealized light pulse creates a quasihole, we will in the following consider different processes which cause decoherence, and which could reduce the fidelity of our scheme: (a) electron-electron interactions, (b) spontaneous emission from the excited level, or  non-radiative dissipation (heating). 

\paragraph{Electron-electron interactions.} Since interaction in the $n=1$ Landau level also support a Laughlin-like phase, interactions will not cause decoherence once the full population has been transferred from one Landau level into the other, that is to say, the initial and the final state will not be affected by interactions. However, during the transfer both Landau levels are occupied, and inter-level interactions differ significantly from intra-level interactions, cf. Ref. \onlinecite{areg}. A straightforward strategy to keep the resulting decoherence small is by using short pulses, that is, by applying a strong Rabi frequency. 

\begin{figure}[t]
\centering
\includegraphics[width=0.48\textwidth, angle=0]{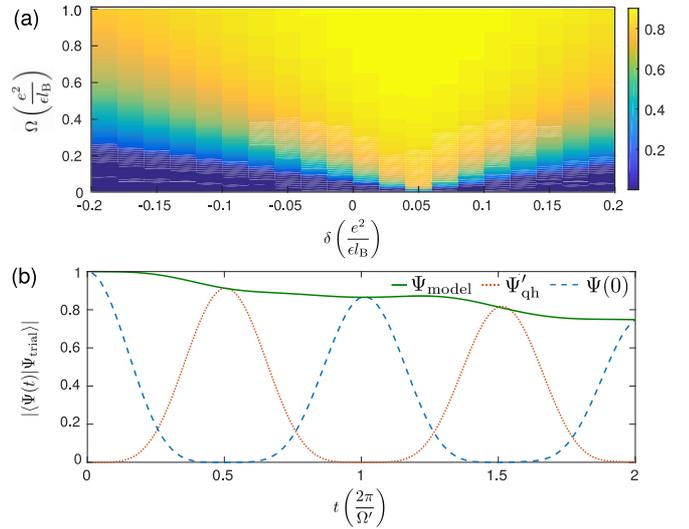}
\caption{\label{fig2} {\bf Fidelity of the quasihole pump.} We consider a system of $N=5$ electrons, initially prepared in the ground state of $V$ within the $n=0$ Landau level at total angular momentum $L_n$. This initial state $\Psi(0)$ has a large ($>0.99$) overlap with the Laughlin state. We then evolve this state  under $H_0+V_{\rm C}$, and consider the overlap of the evolved state $\Psi(t)$ with other trial wave functions, including the one for the quasihole state. In (a), we plot the maximally attained overlap between evolved state $\Psi(t)$ and quasihole state $\Psi_{\rm qh}'$ (defined in Eq. (\ref{qhprime}) as a function of the detuning $\delta$ and the Rabi frequency $\Omega$. In (b), we plot the overlap between $\Psi(t)$ and different trial wave functions as a functions of time. This includes the overlaps with the initial state $\Psi(0)$ (blue dashed line), with the quasihole state $\Psi_{\rm qh}'$ (red dotted line), and with the model wave function $\Psi_{\rm model}(t)$ given in Eq. (\ref{model}) (green solid line). Here we have chosen coupling parameters $\Omega=0.2 e^2/\epsilon l_{\rm B}$ and $\delta=0.04 e^2/\epsilon l_{\rm B}$. Units of time are given as inverts of $\Omega' = \sqrt{\Omega^2+\delta^2}$.
}
\end{figure}

To make this assessment more quantitative, we have numerically simulated the time evolution under $H=H_0+V_{\rm C}$ for $N=5$ electrons, where the single-particle part $H_0$ is defined in Eq. (\ref{H0}), and $V_{\rm C}$ denotes Coulomb interactions. In the simulation, we have restricted the Hilbert space to the two coupled Landau levels, and 
the angular momentum of the initial state fixes the quantum number $\sum_i (m_i-\ell n_i)$. For the initial state $\Psi(0)$, we have chosen the ground state of $V_{\rm C}$ within the $n=0$ level at fixed total angular momentum $L_N$. This state has large overlap ($\sim0.99$) with the Laughlin state. We then determine the overlap of the evolved state $\Psi(t)$ with the state $\Psi_{\rm qh}' \equiv \prod_i a_i^\dagger f_{\rm qh}^0 \Psi(0) $, that is, a state obtained from the initial state by introducing a quasihole and raising the Landau level index of all electrons. In Fig. \ref{fig2}(a), we plot the maximally attained overlap during the course of the evolution as a function of the detuning $\delta$ and Rabi frequency $\Omega$. As a promising result, we find that the Rabi frequency does not have to be much larger than the many-body gap for the fidelity to reach values close to one. 
We note that the many-body gap above the Laughlin phase is on the order $0.15 e^2/\epsilon l_{\rm B}$. This value corresponds to 0.3 eV, if we assume a typical magnetic field strength of 10 teslas, and use the permittivity of the vacuum, $\epsilon=\epsilon_0$.
Our numerical simulation also shows that the best choice for the detuning is not at resonance, but at about  $\delta=0.05~e^2/\epsilon l_{\rm B}$, that is, for an optical frequency below the Landau level resonance. The value of the detuning roughly compensates the interaction energy difference when electrons are pumped into the quasihole state.  Due to a larger total angular momentum in the quasihole state, the Coulomb repulsion in this state is decreased, and the many-body resonance is shifted away from the single-particle value.

\begin{figure}[t]
\centering
\includegraphics[width=0.48\textwidth, angle=0]{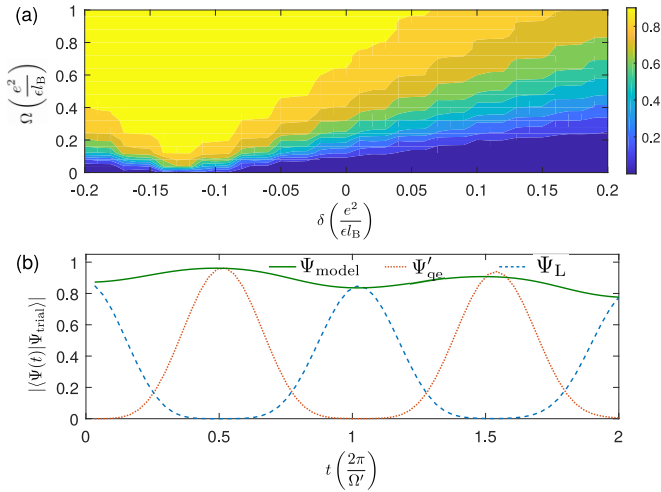}
\caption{\label{fig2qe} {\bf Fidelity of the quasielectron pump.} The setup in this case is similar to Fig.~\ref{fig2}, but the pump photons have orbital angular momentum $l=-1$. Moreover, we simulate pumping in the presence of an additional potential in the lowest Landau level acting on $m=0$ state, to initially remove the population of this state (see main text). (a) We plot the maximally attained overlap with the quasielectron state during a pumping cycle. (b) We plot the overlap of the time-evolved state with the quasielectron state (shifted into LL1) $\Psi_{\rm qe}'$ (red dotted line), the Laughlin state $\Psi_{\rm L}$ (blue dashed line), or a time-dependent model wave function $\Psi_{\rm model}(t)$ (green solid line). The coupling parameters are $\Omega=0.4 e^2/\epsilon l_{\rm B}$ and $\delta=-0.1 e^2/\epsilon l_{\rm B}$}.
\end{figure} 

\paragraph{Spontaneous emission.} Spontaneous emission limits the lifetime of any state above the Fermi level. Therefore, we need to prepare the state of interest in the Landau level at the Fermi energy.  This can be achieved by applying two subsequent $\pi$-pulses, as shown in Fig. \ref{fig:raman}: The first pulse, with orbital angular momentum $\ell=1$, transfers the electrons into an excited Landau level, and  simultaneously shifts the orbital quantum number $m$ to $m+1$, as discussed above. The second pulse with $\ell=0$ returns the electrons to the original Landau level, without changing orbital states. Using sufficiently large Rabi frequencies, both pulses can operate at large fidelities.  The combination of both pulses then results in a quasihole excitation within the Landau level at the Fermi surface. With this scheme, spontaneous emission can only occur during the pulse duration. To neglect this effect, we have to demand that the lifetime in the excited level is large compared to the duration $t=\pi/\Omega$ of a $\pi$-pulse. In other words, 
the coupling has to be fast compared to the emission rate. In summary, large Rabi frequencies (on the order of eV) suppress both decoherence due to interactions or due to spontaneous emission. However, strong Rabi couplings also lead to non-radiative losses. This will set a practical limit to the Rabi strength of the pulse, and thus, to the fidelity of our scheme. A further requirement on the Rabi frequency is that it is large compared to the disorder potential, which ensures that the selection rules for orbital angular momentum are well obeyed.

Within our simulation, we have also studied how the system evolves from the initial Laughlin-like state (polarized in $n=0$ at $t=0$) into a quasihole state (polarized in $n=1$ at $t=\pi/\Omega'$ with $\Omega'\equiv \sqrt{\delta^2+\Omega^2}$). It is found that at intermediate times $0<t<\pi/\Omega'$ the system evolves through a series of edge-like excitations. The most relevant edge states, denoted by $\Psi^{(s)}$, are of the form:
\begin{align}
 \Psi^{(s)} =  \frac{1}{\sqrt{\binom{N}{s}}} \sum_{ \{ k_1,\dots,k_s\} } (-1)^{\sum_{j=1}^s k_j} \prod_{j=1}^s \tilde a_{k_j}^\dagger \tilde b_{k_j}^\dagger \Psi(0).
\end{align}
Here the sum is over all $s$-tuples $\{k_1,\dots,k_s\}$ with $1\leq k_1 <\dots<k_s \leq N$, i.e. all ways of choosing $s$ out of $N$ particles. For $s=0$ and $s=N$, this definition recovers the initial state $\Psi(0)$ and the quasihole state $\Psi_{\rm qh}'$ of Eq. (\ref{qhprime}). 
Generally, $s$ specifies the number of electrons in $n=1$, which is equal to the excess of angular momentum quanta with respect to the Laughlin value $L_N$. Thus, the states $\Psi^{(s)}$ interpolate between edge and quasihole excitations.
By definition, a quasihole excitation increases the total angular momentum by $\mathcal{O}(N)$, while an edge excitation is characterized by an increase of $\mathcal{O}(1)$. 

We also note that the family $\Psi^{(s)}$ contains only a selection of edge states, namely those where $s$ electrons are excited by only one angular momentum quantum.
Other edge states are barely relevant for the dynamics in our system, and we can model with high fidelity the system evolution using only states  $\Psi^{(s)}$ with $0\leq s \leq N$. Therefore, we make the following ansatz:
\begin{align}
\label{model}
 \Psi_{\rm model}(t) = \sum_{s=0}^N 
 {\cal N}_s
 \cos(\frac{1}{2} \Omega' t)^{N-s} \sin(\frac{1}{2} \Omega' t)^s \Psi^{(s)},
\end{align}
where  ${\cal N}_s=(-i)^{{\rm mod}(s,2)}\sqrt{ \binom{N}{s} }$. In Fig. \ref{fig2}(b) we plot, for $N=5$ electrons, the overlap of the exact state with this model wave function as a function of time. Despite our choice of a relatively weak Rabi frequency, $\Omega=0.2 e^2/\epsilon l_{\rm B}$, the model wave function $\Psi_{\rm model}(t)$ captures the evolution rather well.

This establishes the following picture for our quasihole pump: During a pumping period, the quasihole state is reached via a series of edge excitations.  Therefore, as shown by Eq. (\ref{model}), for large systems our scheme requires fine-tuning of the pumping period in order to reach the quasihole state. As a function of time, the overlap with the quasihole state behaves like $\sim\sin(\frac{1}{2} \Omega' t)^N$, so the time window where the evolved state has large overlap with the quasihole state becomes short for large $N$. 

We note that the scenario here is different from another mechanism to produce a quasihole which has been discussed in the context of cold atoms \cite{paredes01}, and which is based on a local repulsive potential. By adiabatically increasing the potential strength, the Laughlin state is turned into the quasihole state without involving significant contributions from other states. In contrast to our scheme, this approach involves a first-order phase transition which would make the adiabatic evolution prohibitively slow in the thermodynamic limit. On the other hand, in small systems, the method provides an intermediate superposition between Laughlin state and quasihole state which can be used for interferometric measurements. 

\begin{figure}[t]
\centering
\includegraphics[width=0.48\textwidth, angle=0]{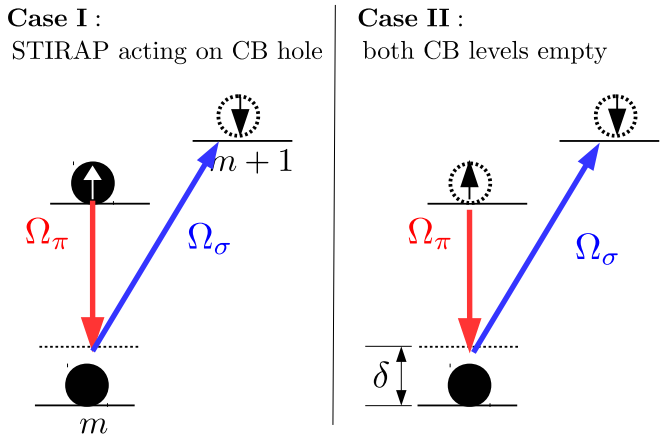}
\caption{\label{cases} {\bf Distinct cases in the STIRAP scheme.} In Case I, the $\uparrow$-electron (black ball) is transferred into the empty $\downarrow$ state (empty dotted ball) via a coupling to the filled valence band (VB) level. Under a particle-hole transformation, this process maps onto a single-particle process, flipping the spin of the conduction band (CB) hole, and the standard single-particle STIRAP scenario applies. In Case II, both CB states are empty, but coupled to a filled VB state. To avoid Rabi oscillation of the valence band electron, both couplings $\Omega_\pi$ and $\Omega_\sigma$ must be sufficiently strongly detuned from the single-photon resonance.
}
\end{figure}

The optical scheme makes it possible to induce the quasielectron state in a similar manner as for the quasihole. To this end, we consider an optical pump with angular momentum $l=-1$ and follow a similar procedure as outlined above. An important difference in this case is that the pump cannot transfer the electron with angular momentum $m=0$ to the higher Landau level due to the absence of the resonantly coupled state. In contrast, in order to produce the quasielectron wave function, the components of the Laughlin wave function for which the $m=0$ state is occupied should be destroyed by the quasielectron prefactor. Given the strongly correlated nature of the Laughlin ground state, this might appear as a challenge for generating the quasielectron state. 
Particularly, in exact diagonalization calculations with finite number of electrons and using the Laughlin state as the initial state, the highest value of the overlap which we were able to obtain with the quasi-electron state is 87\% (not shown).
However, the overlap can be significantly increased to 99\% (Fig.~\ref{fig2qe}), if we add a repulsive potential acting on the $m=0$ state in the lowest Landau level. Such potential excludes the $m=0$ state from the initial wave function, which thus deviates from Laughlin state (Fig.~\ref{fig2qe} (b)). The addition of this potential might appear as a mathematical artifact, but it clearly demonstrates that in the absence of a $m=0$ (as on an annulus), the quasielectron wave function can be produced by our pumping scheme. Moreover, optical methods may even allow to directly implement such a potential in real experiments. 
To this end, one could exploit optical Stark effect (which is discussed in the next section) using an off-resonant coupling from the Fermi level (e.g. $(n=0$) to another Landau level (e.g. $n=-1$). If the light beam has angular momentum $l=1$, it will couple levels $(n=-1,m)$ and $(n=0,m+1)$, and the level $(n=0,m=0)$ remains uncoupled. Thus, the AC Stark effect shifts the energy of all holes in the $n=0$ Landau level, except for the $m=0$ state, and effectively can repel electrons from this state. The large overlap with the quasielectron state which is achieved in exact diagonalization, after we implement such potential, shows the promise of our method for the controlled generation of quasielectrons, and for manipulating such states. As can be seen from Fig.~\ref{fig2qe} (b), the model wave function (\ref{model}) with the appropriate change of $\tilde b_i^\dagger$ with $\tilde b_i$ captures the dynamics quite well for this case as well. The maximum overlap is now attained for detuning $\delta=-0.12~e^2/\epsilon l_{\rm B}$, that is, at opposite sign compared to the quasihole pumping. This is due to the fact that producing a quasielectron reduces the total angular momentum in the system, and thus leads to increased Coulomb repulsion. As a consequence, the many-body resonance is shifted to an optical frequency above the Landau level gap.

\subsection{STIRAP spin-flip scheme in GaAs \label{GaAs}}
We now consider an alternative coupling scheme which is illustrated in Fig. \ref{fig:raman}(b).
The goal and the general strategy is the same as in the previous subsection, but instead of selectively coupling Landau levels, we now achieve the desired momentum transfer via a Raman spin flip process. This approach avoids the need of an anharmonic Landau level spectrum, and thus it can be applied to non-relativistic materials. On the other hand, coupling different spin manifolds in the conduction band to the same level in the valence band requires the presence of spin-orbit coupling \cite{imamoglu00,vanderwal11}. Therefore, the approach in this section is well suited for GaAs, but it cannot be applied to graphene. One of the Raman beams shall carry orbital angular momentum, such that the coupling effectively transfers conduction band electrons from $\ket{n=0,m,\uparrow}$ into $\ket{n=0,m+1,\downarrow}$. As discussed before, the angular momentum transfer creates a quasihole state, but now, as an additional bonus, the final state remains in the $n=0$ Landau level, with only the electron spin being flipped. Given the long spin lifetimes of the order of nanoseconds, measured for GaAs in Refs. \onlinecite{kikkawa98,ohno99}, the final state is effectively stable. Moreover, since the Coulomb interactions are spin-independent, the final state will not be subject to decoherence due to interactions. We remark that our approach is robust against sample disorder provided the timescale for the optically induced transfer process is much faster than the characteristic frequency of the disorder potential.

In order to avoid excitations from the valence band, the timing of the light fields may follow a STIRAP protocol (see Ref. \onlinecite{vitanov17} for a recent review on STIRAP techniques). In the standard STIRAP scenario, a particle is transferred between two stable states, involving two fields which couple these states to a third radiative level. The characteristic feature of STIRAP is the fact that, for properly timed fields, full state transfer is possible without populating the radiative level at any time. 
Our case, though, is different from the standard STIRAP scenario: While we also want to transfer a particle between two (relatively) stable conduction band states, we achieve this via a coupling to a filled valence band level. The scenario is illustrated as Case I in Fig. \ref{cases}. Although this process involves two electrons, STIRAP can be applied if we view the process as the transfer of a single hole. This particle-hole transformation only requires to interchange pump- and probe-fields, and Coulomb interactions between the electrons simply renormalize the resonance frequencies. Our situation, however, is more complicated through the presence of a second scenario, illustrated as Case II in Fig. \ref{cases}. This case may occur whenever the many-body state is at fractional filling, such that the couplings also act onto empty orbitals. This scenario may give rise to undesired excitations from the valence band. A natural way to avoid these excitations is by operating far detuned from a single-photon resonance.

In order to achieve high fidelities, the STIRAP protocol should be characterized by $T \gg 2\pi/\Omega \gtrsim 2\pi/\delta$. Here, $T$ denotes the duration of a STIRAP pulse, which needs to be significantly longer than a $\pi$-pulse. However, due to the spin-independence of Coulomb interactions, there is no need to keep $T$ small compared to the time scale of interactions, set by $e^2/(\epsilon l_{\rm B})$, where $\epsilon=\epsilon_0\epsilon_{\rm d}$. Note that in GaAs, a dielectric constant $\epsilon_{\rm d}\approx 12$ suppresses Coulomb interactions by a factor of 12 compared to graphene. Thus, at a field strength of about $B\approx10 \ {\rm T}$, the energy scale of Coulomb interactions is on the order of tens of THz, i.e., at femtosecond time scales. In contrast, the lifetimes of spin excitations is on the order of nanoseconds, so it is justified to treat both spin levels as stable levels.

\begin{figure*}[t]
\centering
\includegraphics[width=0.88\textwidth, angle=0]{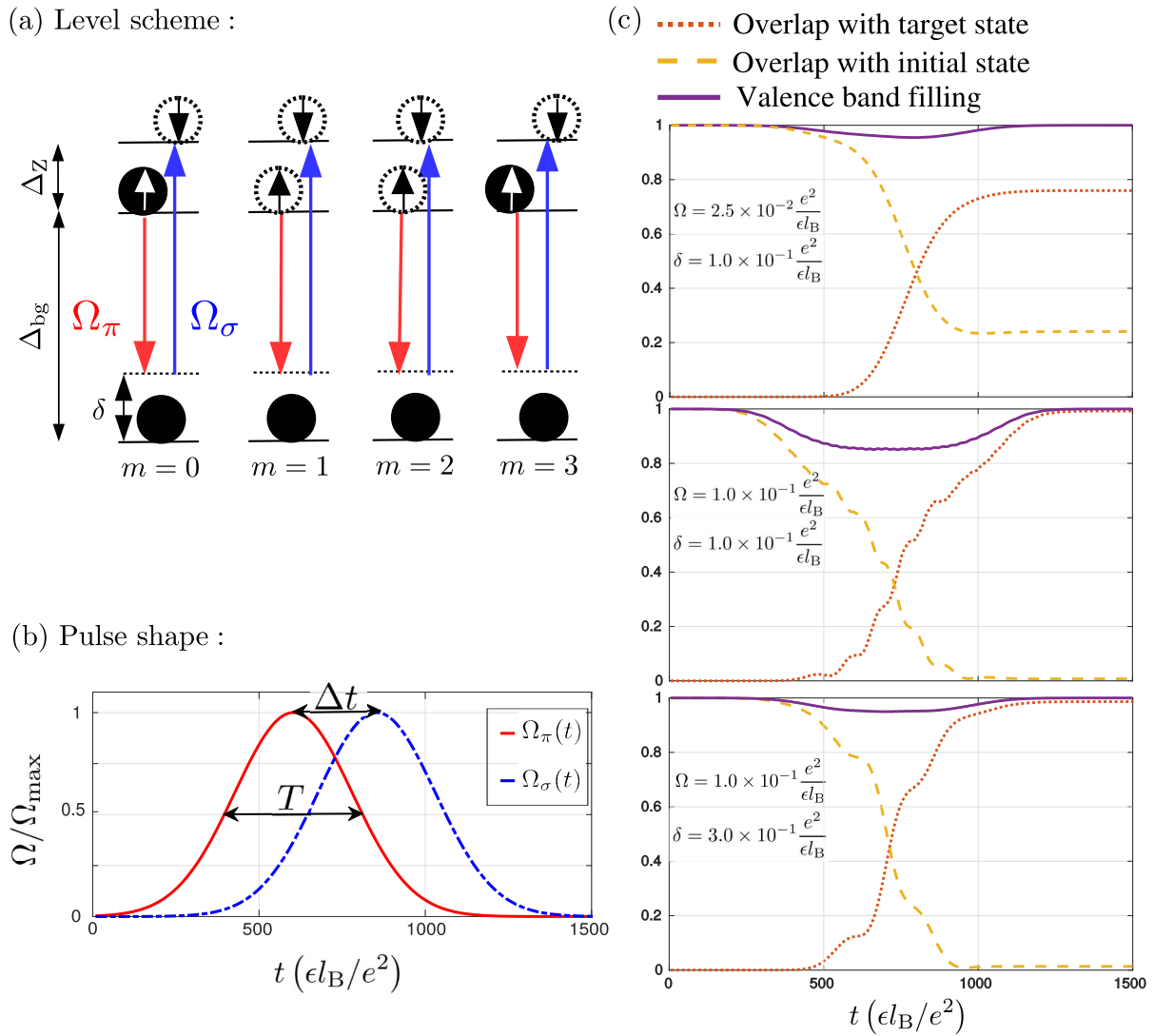}
\caption{\label{stirap} {\bf Simulation of many-body STIRAP scheme.} 
(a) Illustration of the STIRAP scheme for which we have performed a numerical simulation in a minimal system: Each Landau level is truncated at $m=3$ (4 states), and the total number of electrons is six. For simplicity, we have considered a coupling at constant angular momentum, i.e. between $\ket{m,\uparrow}$ and $\ket{m,\downarrow}$. Bandgap $\Delta_{\rm bg}$ and Zeeman gap $\Delta_{Z}$ are chosen such that in the absence of the coupling, 4 electrons fill the valence band, and the remaining 2 electrons polarize in the spin-up manifold of the conduction band, where they form a $N=2$ ``Laughlin'' state, $\Psi\sim(z_1-z_2)^3$. (b) Applied STIRAP pulses, with the pulse duration $T=415 \epsilon l_{\rm B}/e^2$, that is, on the order of hundreds of picoseconds. The pulse maxima are separated by $\Delta t=250\epsilon l_{\rm B}/e^2$. (c) We simulate the time evolution for different values of Rabi frequency $\Omega$ and detuning $\delta$, obtaining the valence band filling (occupation per states), and the overlap with initial state (a $N=2$ Laughlin state in the spin-up manifold) 
and target state ((a $N=2$ Laughlin state in the spin-down manifold), as a function of time. In the upper plot of panel (c), the transfer is poor due to a relatively weak Rabi frequency $\Omega=0.025 e^2/(\epsilon l_{\rm B})$, and comparably strong detuning $\delta=0.1 e^2/(\epsilon l_{\rm B})$. The plot in the center is for an increased Rabi frequency $\Omega=\delta=0.1e^2/(\epsilon l_{\rm B})$, which allows for good transfer, but excitations from the valence band become relatively large. The lower plot, for $\Omega=0.1e^2/(\epsilon l_{\rm B})$ and $\delta=0.3e^2/(\epsilon l_{\rm B})$, leads to good transfer at low excitation rates.
}
\end{figure*}

Since the transfer dynamics now involve three different Landau levels, including the filled valence band Landau level, its numerical simulation is hard. We restrict ourselves to simulating a minimal example of the scheme: As plotted in Fig. \ref{stirap}(a), we truncate the Landau levels to having only 4 states, and load the system with $N=6$ electrons. Choosing the bandgap and the Zeeman gap sufficiently large compared to the Coulomb interactions, the ground state of this system will be a filled valence band, and two electrons in the spin-up manifold of the conduction band, forming a ``Laughlin'' state of two electrons, i.e. $\Psi(z_1,z_2) \sim (z_1-z_2)^3 \ket{\uparrow \uparrow}$. If we wanted to pierce a hole into this state, this would increase the angular momentum by 2, which is not permitted in our truncated Landau levels. Therefore, we simulate only the coherent transfer part of our scheme, that is, a coupling as shown in Fig. \ref{stirap}(a), leaving the angular momentum constant. Applying the STIRAP protocol shown in Fig. \ref{stirap}(b), we evaluate the fidelity, that is, the overlap of the time-evolved state $\Psi(t)$ with the target state,  $\Psi_{\rm target}(z_1,z_2) \sim (z_1-z_2)^3 \ket{\downarrow \downarrow}$. As seen in Fig. \ref{stirap}(c), this fidelity reaches unity, if the Rabi frequency is sufficiently strong, i.e. $\Omega T \gg 2\pi$. For the pulse duration (defined by the FWHM of a Gaussian pulse), we have chosen $T\approx 415 \epsilon l_{\rm B}/e^2$, which for typical magnetic field strengths is on the order of hundreds of picoseconds. To avoid excitations from the valence band, the detuning should be larger than the Rabi frequency, $\delta > \Omega$.

\subsection{Detecting anyonic properties \label{sec:corbino}}
In the remainder of this section, we briefly discuss possible detection schemes for fractional charge and statistics, which potentially benefit from a method to generate exactly one quasihole by a pulsed light beam.

\paragraph{Fractional charge.}
A possible charge measurement can be performed on a Corbino disk. The insertion of flux through a Corbino disk has been discussed in Ref. \onlinecite{thouless90}. As described in the previous section, our scheme increases the angular momentum of the electrons. This shifts them towards the outer edge in the same way as creation of an additional flux through the inner circle of the Corbino disk would do. The reverse process, which transports charges towards the inner edge can be achieved by decreasing the angular momentum of the electrons. 

The confining potential at the edge makes it energetically favorable for the charge to return to its original position. This leads to transport through a wire connecting the two edges of the Corbino disk. However, considering a fractional quantum Hall system at filling $\nu=1/q$, $q$ quasiparticles need to be shifted to the outer edge in order to accumulate a total electronic charge $e$. Thus, $n$ pumping cycles are expected to produce a current of $n/q$ electrons, and the number of pumping cycles serves as a direct measure of the fractional charge.

\paragraph{Fractional statistics.}
The detection of fractional statistics is possible using interferometers, either of the Fabry-Perot or the Mach-Zehnder type. Such devices are suited to detect Aharonov-Bohm phases, as proposed for instance in Ref. \onlinecite{chamon97} and realized in Refs. \onlinecite{camino05,mcclure12,willett13}, by measuring the interference of currents along different paths. In these schemes, the interference pattern is sensitive to changes of the magnetic field, which yields the value of the fractional charge. It is also sensitive to the number of quasiparticle between the different arms of the interferometer, and from this, the statistical angle of the excitations can be deduced. However, to extract both charge and statistical angle from the interference pattern, exact knowledge about the number of excitations is needed. Thus, our scheme may allow for improved measurements as it provides individual control over these excitations.

\section{Light-induced potentials:}
\label{pin}
The previous section has demonstrated that light with orbital angular momentum can be used to mimic the addition of a flux, and to produce a quasiparticle excitation. In the present section, we will not be concerned with the production of the excitation, but we will be interested in ways to stabilize and control the quasiparticle. Specifically, we will 
consider an optical potential which locally repels the electrons and thereby traps a quasihole. 

A major concern addressed in this section is the finite width of the optical potential, in contrast to $\delta$-like potentials which have been studied earlier in the context of cold atoms \cite{paredes01,julia-diaz12}. A numerical investigation shows that a potential with small but finite width is even better suited for trapping quasiholes than a point-like potential. However, the gap above the quasihole state is found to decrease when the potential becomes broader than the magnetic length. Since the optical wavelength is usually larger than the magnetic length, we will present some ideas to achieve subwavelength potentials using a three-level coupling.

Given the flexibility of optical potentials, they appear to be particularly well suited for moving the quasihole. Thus, an optical trap for quasiholes may provide a tool for braiding anyonic excitations. To demonstrate that ability, we show that, when the potential is moved on a closed contour, the wave function acquires a Berry phase proportional to the fractional charge of the quasihole. 

The calculations and discussions in this section hold for both non-relativistic systems and for Dirac materials.

\subsection{AC Stark shift}
The mechanism which provides the desired optical potential is AC Stark shift. The AC Stark shift is routinely used to trap cold atoms in optical lattices. Recently, it has been suggested to trap Dirac electrons in graphene by exploiting AC Stark shift \cite{morina18}. In a GaAs quantum well, this shift can be produced by optically pumping below the band gap \cite{vonlehmen86}. Alternatively, if the system is coupled to a cavity, an enhanced Stark shift can be achieved using a resonance of the cavity \cite{edo}. In general, the energy shift $\Delta E$ experienced by the electronic energy in a laser field ${\bf E}({\bf r},t)$, is given by $\Delta E= {\bf d}\cdot {\bf E}({\bf r},t)$, where ${\bf d}= \alpha [ E_x({\bf r},t),E_y({\bf r},t)]$ is the dipole moment induced by the field. The polarizability $\alpha$ is inversely proportional to the detuning $\Delta$ from the closest resonance. With this, the optical potential reads:
\begin{align}
 V_{\rm opt} \propto \frac{I(z)}{\Delta},
\end{align}
where $I(z)$ is the laser intensity in the complex plane, assumed to be constant in time. In the following, we will consider a Gaussian beam, that is, an optical potential $V_{\rm opt}^{(\xi,w)}(z) = \left(\frac{l_{\rm B}}{w}\right)^2 V_{{\rm opt},0} \exp[|z-\xi|^2/w^2]$, characterized by the position of the beam focus $\xi$, the width $w$ of the beam, and an potential strength $V_0$. The prefactor $ \left(\frac{l_{\rm B}}{w}\right)^2$ normalizes the intensity such that $\lim_{w\rightarrow 0} V_{\rm opt}^{(\xi,w)}(z)= V_{{\rm opt},0} \delta(z)$, with $\delta(z)$ being the Dirac distribution. 

 For the purpose of trapping a quasihole, it is necessary that the strength of the potential compensates the energy gap above the Laughlin state. Thus, the relevant energy scale is given by the Coulomb energy $e^2/(\epsilon l_{\rm B})$, with the magnetic length $l_{\rm B}$ representing the typical length scale relevant for the quantum Hall physics.
This length scale determines the size of an electronic orbital, but also of defects like  quasiparticles and quasiholes.  If $\tilde B$ is the magnetic field strength in tesla, $l_{\rm B} = 26 \ {\rm nm}/\sqrt{\tilde B}$. For a magnetic field of 9 T, and with a dielectric constant $\epsilon_{\rm d}=12$ (as in GaAs), this energy scale is on the order of 150 meV, and a typical gap will be on the order of 15 meV. An early measurement in GaAs \cite{vonlehmen86} has obtained an AC shift of 0.2 meV was with a laser intensity of 8 $\rm MW/cm^2$. 

Apart from the energy scale, also the length scale of the potential plays an important role. With the size of a quasihole being on the order of the magnetic length, we expect that the length scale of a trapping potential should not significantly exceed this scale. However, the minimum length scale of an optical potential is limited by the wavelength of the light, which in the visible regime is on the order of hundreds of nanometers. In contrast, for magnetic field strengths on the order of a few teslas, the magnetic length is only a few nanometers. We will thus need to evaluate whether a potential with finite width $w \gg l_{\rm B}$ is still suited to trap quasiholes.
 
 \subsection{$\delta$-like potentials}
 Before considering the case of finite-width potentials, we verify that a point-like potential ($w=0$) gives rise to the desired excitations. This becomes obvious when we look at the parent Hamiltonian of the Laughlin state, that is, at some model interactions $V_{\rm parent}$ for which the Laughlin wave function $\Psi_{\rm L}$ is the densest  zero-energy eigenstate. Such parent Hamiltonian is given in terms of Haldane pseudopotentials $V_m$, specifying the interaction strength between two electrons at relative angular momentum $\hbar m$. In the 1/3-Laughlin state, all pairs of electrons have relative angular momentum $3\hbar$, so the Laughlin state has zero energy in a model Hamiltonian with $V_m=0$ for $m\geq 3$. Since for spin-polarized fermions the relative angular momentum cannot be even, a Hamiltonian with only a single non-zero pseudopotential, $V_1$, provides a  parent Hamiltonian for the Laughlin state. It follows that the quasihole state $f_{\rm qh}^\xi \Psi_{\rm L}$ becomes the densest zero-energy eigenstate of $V_{\rm parent}+V_0\delta(\xi)$, when the potential strength exceeds a critical value. To see this, we note that the quasihole state carries the same anti-correlations between the electrons as the Laughlin state, but at the same time has vanishing density at position $\xi$. 
 
 The scenario of a $\delta$-potential has been studied before in greater detail in the context of cold atoms \cite{paredes01,julia-diaz12}. In these systems of neutral particles, which can be brought into the fractional quantum Hall regime by artificial gauge fields, the cyclotron frequency is on the order of the trap frequency ($\sim 10$ Hz), resulting in a magnetic length $l_{\rm B} = \sqrt{\hbar/(M\omega_{\rm c})}$ on the order of microns, with $M$ being the mass of the atoms. Due to this different length scale, finite-width effects can indeed be neglected in these artificial systems. 

 \subsection{Finite-width potentials}
To study the role of the finite potential width $w$, we turn to numerical diagonalization methods, by which we obtain the ground state of $V_{\rm C}+V_{\rm opt}^{(\xi,w)}$ for different laser positions $\xi$ and different beam widths $w$. Generally, we find large overlaps of these states with the Laughlin quasihole state $f_{\rm qh}^\xi \Psi_{\rm L}$, even when the beam becomes as broad as (or even broader than) the electronic cloud. In our numerics, we have considered both disk and torus geometries which we discuss separately below.

\begin{figure}[t]
\centering
\includegraphics[width=0.48\textwidth, angle=0]{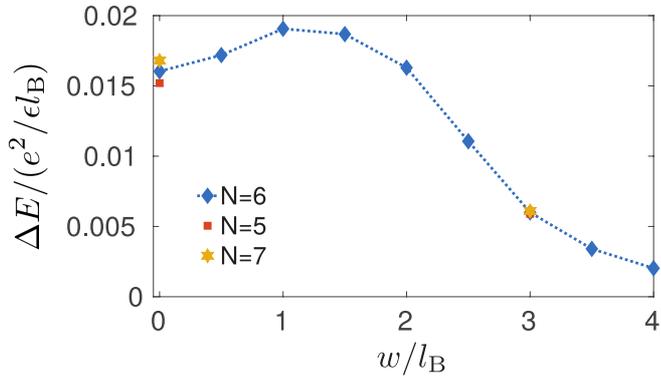}
\caption{\label{gap} {\bf Gap above the quasihole state on a torus.} We plot the energy gap above the three degenerate quasihole states on a square torus in the presence of an optical potential $V_{\rm opt}^{(\xi,w)}$, as a function of the potential width $w$. The $N$ electrons are confined in the $n=0$ Landau level, generated by the presence of $N_{\Phi}=3N+1$ magnetic fluxes.}
\end{figure}

\paragraph{Torus.} The torus geometry is convenient because due to its compact nature no trapping potential is required to confine the electrons. Since the torus has no edge, this geometry is well suited for studying the bulk behavior of large systems for which deformations at the edge are irrelevant. Interestingly, on the torus, the ground state of $V_{\rm C}+V_{\rm opt}^{(\xi,w)}$ is almost independent of $w$. The overlap with the exact Laughlin quasihole state \footnote{The Laughlin quasihole states can be obtained as zero-energy states from the parent Hamiltonian, that is, from pseudopotential interactions with $V_1$ being the only non-zero term, together with a $\delta$-like potential to create the quasihole. On the torus, one needs to consider the Hilbert space of $N$ electrons and $N_{\Phi}=3N+1$ magnetic fluxes, which corresponds to LL filling 1/3 plus one additional flux.} takes large values close to 1, cf. Table \ref{table1}. While this result shows that the finite potential width does not modify the quasihole state in the bulk, we also find that the energy gap above the quashihole states is quite sensitive to the width of the beam (see Fig. \ref{gap}). Up to a certain value, of the order of the magnetic length, a finite potential width is found to increase the stability of the quasihole. However, the gap starts to decrease when the beam width exceeds the magnetic length. 
This result can be understood by noticing that also the quasihole has a finite size of the order of the magnetic length, and the formation of a quasihole reduces the energy due to the optical potential most efficiently when the spatial overlap between quasihole and potential becomes largest. Obviously, in broader potentials a quasihole becomes less efficient for reducing potential energy.

\paragraph{Disk.}
The disk, though the most natural geometry to study quantum Hall physics, suffers strongly from finite-size effects. Even the concept of a filling factor is not defined on an infinite disk because each Landau level contains an infinite amount of states. It is necessary to assume a trapping potential which controls the electron density. A realistic trapping potential consists of hard walls, so the potential is flat in the entire system, except for the edge, where the potential energy steeply increases. Effectively, such potential puts a constraint on the Hilbert space, as it restricts the orbitals to those which fit into the flat region. This means that angular momenta beyond a certain value are not available anymore. Since we are interested in the Laughlin state (with angular momentum $L_N$), and in its quasihole excitation (increasing the angular momentum by up to $N$ quanta), we will assume that the trap effectively truncates the Hilbert space at $L_N+N$. Therefore, we perform the exact diagonalization study within a space of Fock states of angular momentum $L_N \leq L \leq L_N+N$. This choice yields the quasihole state as the only zero-energy eigenstate if the parent Hamiltonian is applied, that is, for a point-like potential $V_{\rm opt}^{(\xi,w=0)}$ and pseudopotential interactions $V_m \sim \delta_{1,m}$. 

\begin{table}
\begin{tabular}{c|c|c}
 $w/l_{\rm B}$ & Overlap on torus & Overlap on disk \\
 & $N=7$ & $N=8$ \\
 & $N_{\phi}=22$ & $84\leq L/\hbar \leq 92$ \\ 
 \hline
 0 & 0.9884 & 0.9450 \\
 3 & 0.9885 & 0.9552 \\
 6 & 0.9851 & 0.9409
\end{tabular} 
\caption{\label{table1}
Overlaps between Laughlin quasihole state, and ground state of $V_{\rm C}+V_{\rm opt}^{(\xi,w)}$ on disk and square torus, for different $w$.
Parameters on the torus: $V_0=1,\xi=0$, and on the disk: $V_0=10,\xi=2$. On the torus, exhibiting three (quasi)-degenerate ground states, overlap refers to the three (equal) eigenvalues of the 3x3 overlap matrix.
}
\end{table}

Importantly, we find that Coulomb interactions do not significantly alter the scenario. Comparing the exact Laughlin quasihole state and the ground state of $V_{\rm C}+V_{\rm opt}^{(\xi,w)}$, we obtain  an overlap of about 0.95 for $N=8$. Strikingly, the potential width $w$ has only a minor effect on these numbers if the potential is chosen sufficiently strong, see Table \ref{table1}.

\begin{figure}[t]
\centering
\includegraphics[width=0.48\textwidth, angle=0]{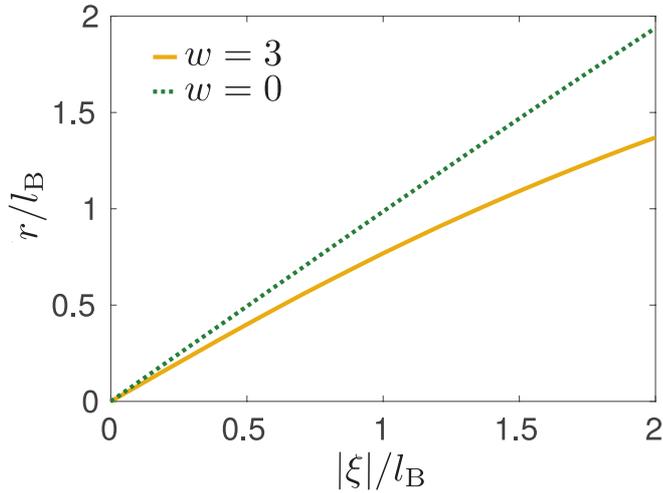}
\caption{\label{rxi} {\bf Calibrating the quasihole position on the disk.} By neglecting the trapping energy, long-range interactions lead to a shift of the radial position $r$ of the quasihole towards the center.  The plotted curve is used to calibrate the true quasihole position $r$ as a function of the parameter $|\xi|$, specifying the maximum of the optical potential, for $N=8$ electrons.}
\end{figure}

There is, however, a notable consequence of finite-range interactions appearing in our numerics on the disk: the quasihole position does not exactly coincide with the position of the optical potential anymore, as seen in Fig. \ref{rxi}. Although this observation seems to be an artifact due to the neglection of the trap, it will be important to take it into account  when determining the quasihole charge, as discussed in the next subsection. To this end, the data in Fig. \ref{rxi} is needed to calibrate the quasihole position.

Energetic arguments explain the mismatch between quasihole position and potential minimum: shifting the quasihole towards the center increases the angular momentum, and thereby reduces the energy of long-ranged interactions. A realistic trapping potential would compensate this effect by penalizing the angular momentum increase, but this term is missing in our numerical study. If Coulomb interactions are replaced by the short-ranged pseudopotential model, the quasihole position coincides with the potential minimum.  In this case, the interaction energy is zero, and shifting the quasihole cannot lead to an interaction energy gain.

\subsection{Realization of \subwavelength potentials}
\begin{figure}[h]
\centering
\includegraphics[width=0.48\textwidth, angle=0]{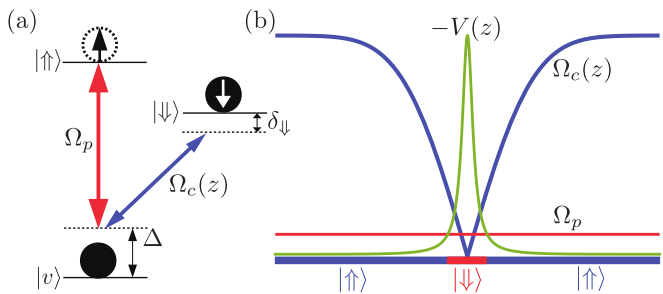}
\caption{\label{fig:sub} 
\textbf{Sub-diffraction potentials.}
(a) 
Engineering a \subwavelength potential via three-level coupling.
An $\Uparrow$-hole at the Fermi level (empty dotted ball) experiences an attractive potential by coupling to the electrons (black balls) in
the $\downarrow$-level of the conduction band and in the valence band state $\ket{v}$. A particle-hole transformation relates this process to the standard single-particle EIT scenario, applied to a single hole. Although the laser fields do not induce a direct potential for $\upSpin$-electrons, the attractive potential for  $\Uparrow$-holes results in an effective repulsive potential for the  $\upSpin$-electrons.
(b) 
We show a 1D cut through the potential $V(z)$ and the laser fields $\Omega_c(z)$ and  $\Omega_p$. 
Even though $\Omega_c$ is diffraction limited, we can achieve a \subwavelength $V(z)$ by working with ${\rm max}[\Omega_c(z)]\gg \Omega_p$.
}
\end{figure}
In the previous section, we showed that the manipulation of anyons profits from potentials of width $w\sim l_B$. This requires a \subwavelength addressability which can be achieved by employing techniques analogous to the ones used in ultra-cold atoms~\cite{Gorshkov2008a,Wang2018}. 
The basic idea is to use three energy levels which provides much more flexibility than just the two-level scheme used to induce AC Stark shift. 
As an example we consider GaAs, for which  we can use the level scheme shown in Fig. \ref{fig:sub}~(a), in analogy to Fig. \ref{cases} used for the STIRAP. The scheme consists of two spin-levels in the conduction band and one level in the valence band. We now choose the Fermi energy through the upper spin level ${\upSpin}$, so both the $\downSpin$-level and the valence band are occupied. The two-electron system can be mapped onto a single-particle problem via particle-hole transformation, and a repulsive potential on ${\upSpin}$-electrons will be achieved by engineering an attractive potential for $\Uparrow$-holes. Therefore, we operate  at the two-photon detuning $\deltaDown<0$. 

Moreover, we use two laser fields: a strong $\OmegaC(z)$ 
which is position dependent [for the easiness of presentation we fix it to $\OmegaC(z)^2=\Omega_0^2(1-\exp[|z-\xi|^2/w^2])$], 
and a weaker $\OmegaP$ which is homogeneous in space. 
  The Hamiltonian reads
\begin{equation}
H_{\rs al}=
\left(
\begin{array}{ccc}
\deltaDown& 0 & \Omega _c(z) \\
0&0& \Omega_p  \\
 \Omega _c(z) & \Omega_p  & \Delta\\
\end{array}
\right)\label{HaEIT}
\end{equation}
in the bare hole-state-basis: $\{\ket{\down},\ket{\up}, \ket{v}\}$.
For $|\deltaDown| \ll\OmegaP$ which ensures that we can consider $\deltaDown$ perturbatively, and for  an appropriate preparation scheme~\cite{Wang2018,Lacki2016}, the internal state of a hole 
can be described using a dark state $\ket{D}=\frac{\OmegaC(z)}{\sqrt{\OmegaP^2+\OmegaC(z)^2}}\ket{\up}-\frac{\OmegaP}{\sqrt{\OmegaP^2+\OmegaC(z)^2}}\ket{\down}$.
From the form of $\ket{D}$, we see that the hole experiences an attractive potential $V(z)=\delta_\down\frac{\OmegaP^2}{\OmegaP^2+\OmegaC(z)^2}$, which for $\Omega_0\gg\OmegaP$ can have \subwavelength width $w_s=w/s$ characterized by the enhancement factor $s\sim \Omega_0/\OmegaP$ and the depth $V_0=\deltaDown$.

Assuming that we can describe our system using three levels,
the available depth of the trap is mainly limited by: (i)  the validity of the rotating wave approximation, and (ii) the coupling to the short-lived intermediate level.
The (i) limitation constrains the strength of $\Omega_c(z)$  to $\Omega_0\ll\Delta{\rs bg}$.
Together with $\Omega_p\gg V_0$ and $s=\Omega_0/\Omega_p$, we get that $s\ll \Omega_0/V_0\ll \Delta_{\rs bg}/V_0$. 
For $\Delta_{\rs bg}\sim 1.5$~eV, $\Omega_0\sim 0.5$~eV,  and $V_0\sim15$~meV, we see that enhancement  factors $s$ on the order of $10$ are within a reach.
The losses in (ii), lead to the broadening of the trapping potential by $\gamma_v \frac{V_0^2}{\Omega_p^2}\ll \gamma_v$, which [compared with the depth of the potential $V_0$] is negligible for the lifetimes $\tau_v=1/\gamma_v$ on the order of  10~ps.
Note that in contrast to ultra-cold atoms~\cite{Lacki2016,Jendrzejewski2016,Wang2018,BieniasInPreparation}, the kinetic energy is quenched in a magnetic field, and therefore non-adiabatic corrections to the Born-Oppenheimer potentials~\cite{Lacki2016,Jendrzejewski2016} are negligible. This relaxes some of the constraints posed on the possible trapping depths. 
We leave a more detailed analysis, beyond the estimates presented here, for the future work.

Finally, in the case of graphene, we envision similar possibilities: for example, one can use other filled Landau levels as the additional two levels in the ladder or lambda three-level scheme.

\subsection{Moving a quasihole.}
In the remaining part of this section, we consider an optical potential which is moved on a closed loop. As a quasihole is trapped by the potential, this procedure is expected to imprint a Berry phase onto the wave function which is proportional to the charge of the quasihole. Thus, by calculating the quasihole charge from the Berry phase we will verify that moving the optical potential is suited to move an excitation. By considering short-range interactions instead of Coulomb interactions, finite-size effects become small, and the fractional charge matches with the value 1/3, expected for a thermodynamically large Laughlin system. We will also compare an idealized adiabatic evolution, restricted to the ground state Hilbert space, with the true dynamic evolution. This establishes the maximal speed with which the potential should be moved.

\paragraph{Relation between Berry phase and charge.} If the position $\xi$ of a single charge $q$ is moved, the wave function $\Psi(\xi)$ will pick up a Berry phase $\gamma=\oint {\rm d}\xi \langle \Psi(\xi) | \bigtriangledown_\xi | \Psi(\xi) \rangle$, and this phase is proportional to the magnetic flux through the enclosed area $A$ times the value $q$ of the electric charge. This relation is normalized such that the electron charge $e$ acquires a Berry phase $\gamma=2\pi$ when it encircles one flux. In a constant magnetic field, with the magnetic length defined such that an area $2 \pi l_{\rm B}^2$ contains one flux quantum, we have the relation
\begin{align}
\label{qe}
 \frac{q}{e} = \gamma\frac{l_{\rm B}^2}{A}.
\end{align}
Thus, by studying the phase of the wave function upon moving the quasihole, we can extract the electric charge of the excitation. 

\begin{figure}[t]
\centering
\includegraphics[width=0.48\textwidth, angle=0]{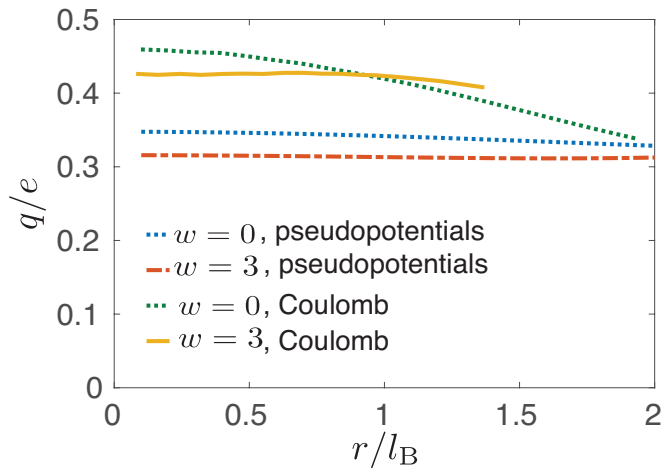}
\caption{\label{charge} {\bf Charge of Laughlin quasihole.}
We plot the charge $q$ of a quasihole in a system of $N=8$ electrons on the disk as a function of radial quasihole position $r$. The total angular momentum is restricted to the Laughlin regime, $L_N\leq L \leq L_N+N$, and the Hamiltonian consists of interactions $V_{\rm int}$ and an optical potential $V_{\rm opt}^{(\xi,w)}$, of width $w$ and focused at position $\xi$.
We have tuned the radial position $|\xi|/l_{\rm B}$ of the optical potential from 0.1 to 2, and obtained the corresponding radial position position $r$ of the quasihole. The potential is then moved on a circle around the origin, which leads to a Berry phase which we evaluate using the static method of Eq. (\ref{gamma_static}) for 200 discrete steps. We consider both Coulomb and pseudopotential interactions, the latter providing a parent Hamiltonian for the Laughlin state. We compare point-like potentials ($w=0$) and finite-width potentials ($w=3l_{\rm B}$). Independently of the potential width, the system with pseudopotential interactions agrees well with the expected value $q/e=1/3$, whereas finite-size effects spoil the numerical value in the system with Coulomb interactions.
}
\end{figure}

\paragraph{Results from adiabatic evolution.}
We have performed such calculation using the disk geometry with $N=8$ electrons. We considered a Hamiltonian $H=V_{\rm int}+V_{\rm opt}^{(\xi,w)}$, where the interactions $V_{\rm int}$ are either Coulomb interactions or the parent Hamiltonian of the Laughlin state. For the optical potential $V_{\rm opt}^{(\xi,w)}$, we considered a finite width $w$ up to $3l_{\rm B}$ as well as the limit $w\rightarrow 0$. Our results are plotted in Fig. \ref{charge}: Importantly, in case of the ideal interactions from the pseudopotential model, the width of the optical potential has a minor effect on the Berry phase, or, respectively the measured charge of the quasihole. Both, for a point-like potential and for a broad beam ($w=3 l_{\rm B}$),  the quasihole charge is almost independent of the quasihole position, as it should be in a quantum liquid. Moreover, the value of the charge is close to the expected value $1/3$. The accuracy of this result despite the small system size is due to the particular choice of interactions. As the pseudopotential model has short range interactions, finite-size effects are much weaker than in the long-range Coulomb case. For Coulomb interactions, the charge as a function of quasihole position is found to be $>0.4e$, that is, it differs significantly from $e$/3. Surprisingly, in the presence of Coulomb interactions, the results for the finite-width potential are closer to the ideal value 1/3.

The important conclusion which we draw from Fig. \ref{charge} is that the finite width of the optical potential does not appear as a limiting factor for a charge measurement by moving the potential. In other words, the observed behavior suggests that even if the optical trap is much wider than the actual size of a quasihole, the quasihole still follows the contour described by the moving potential, and despite their broad width optical potentials can be used for moving and braiding anyons.

The results shown in Fig. \ref{charge} were obtained from an ``adiabatic'' simulation, that is, we actually did not simulate the dynamics of the system while the potential is moved, but we assumed that for any potential position the system remains in its ground state. Thus, we simply determine the ground state $\Psi_n$ at different quasihole positions $R e^{i\varphi_n}= R e^{i n \Delta \varphi}$, and obtain the phase difference between subsequent states from their overlap. Summing up all phase differences along the contour gives the Berry phase
\begin{align}
\label{gamma_static}
\gamma =  \sum_n {\rm Im}\left( \langle \Psi_{n+1} | \Psi_n \rangle -1\right).
\end{align}
In this approach, it is important to fix the global gauge.  In our case of a non-degenerate ground state, the possible global gauge transformations are $U(1)$ rotation. Since we compare eigenvectors obtained from two different diagonalization procedures, we have to assure that the global gauge remains the same.  This can be done by demanding that a certain component of the state vector is real and positive \cite{kohmoto}. However, this procedure requires that some assumptions and conditions hold: Of course, any state along the path needs to have a non-zero overlap with this reference state.  Moreover, one must ensure that, after discretization of the parameter space, the global gauge transformation does not remove the Berry phase. We can achieve this by choosing a reference component which does not gain a phase when the potential is moved.  It is easy to find such a component, since we know that the quasihole is described by a wave function of the type $\prod_i (z_i-\xi)\Psi$. This means that the part of the wave function given by $\prod_i z_i\Psi$ is not affected by the quasihole position. Any component which has non-zero overlap with this expression can be used as a reference component, that is, any occupied Fock state with angular momentum $L=L_N+N\hbar$.

\paragraph{Dynamical evolution.}
In the remainder, we compare the ``adiabatic'' approach with a dynamic one. In the dynamic approach, we really simulate the time evolution of the system while the potential is moved, without assuming adiabaticity of the process. Of course, the dynamic approach is much more costly, as it requires full diagonalization of the Hamiltonian, whereas in the static approach only the ground state is needed. But there are some conceptual advantages of the dynamic method: First, this method yields the overlap between initial and final state which provides a measure for the adiabaticity of the process. From this one can also obtain information about how fast the optical potential may be varied. Second, the dynamic method does not require the gauge fixing procedure described above.

For our dynamical simulation we discretize time, and define $H_n$ as the Hamiltonian with the optical potential $V_{\rm opt}$ at position $R e^{i\varphi_n}$. We then evolve for short periods $\Delta t$ under $H_n$, applying the evolution operator $U_n=\exp\left(i H_n \Delta t\right)$ to the quantum state, and afterwards we quench from $H_n$ to $H_{n+1}$. 
 Starting from $\Psi_0$, the ground state of $H_0$, we reach a final state $\Psi=\prod_{n=1}^{n_{\rm max}} U_n \Psi_0$, where $n_{\rm max}=2\pi/\Delta \varphi-1$. If the process was adiabatic, initial and final state only differ by a phase $\langle \Psi | \Psi_0 \rangle = e^{i \phi}$. This phase now consists of a dynamical contribution $\phi_T$,  determined by the energy of the state, and the Berry phase $\gamma$. In the particular case of a circular rotation around the origin, the energy does not change, and $\phi_T=E_0 T$ where $T=n_{\rm max}\Delta t$. Thus, the Berry phase is obtained by
\begin{align}
 \label{gamma_dyn}
 \gamma= {\rm Im}\left({\rm ln}  \langle \Psi | \Psi_0 \rangle \right)-E_0 n_{\rm max}\Delta t.
\end{align}

\begin{figure}[t]
\centering
\includegraphics[width=0.48\textwidth, angle=0]{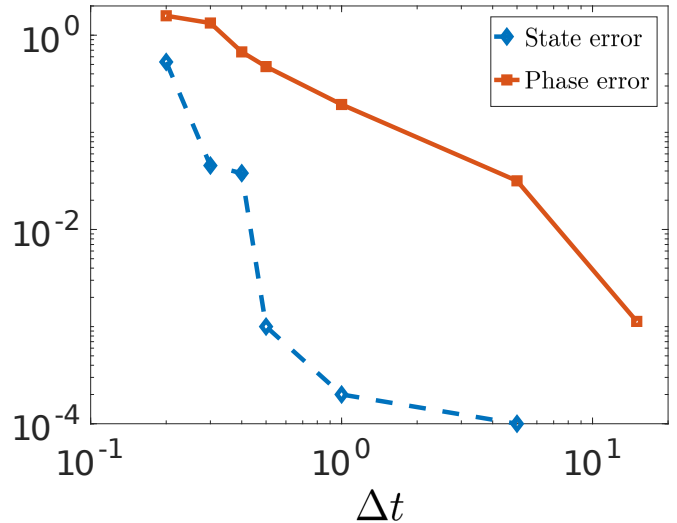}
\caption{\label{adi} {\bf Deviations from adiabatic process due to finite times.}
We compare two types of errors occurring if the quasihole is not moved adiabatically, as a function of the time step duration $\Delta t$ (in units $\hbar \epsilon l_{\rm B}/e^2$).
The red curve shows the relative phase error, defined as $\Delta \gamma/\gamma_{\rm ad}$. Here, $\Delta \gamma$ is the difference between the adiabatically obtained Berry phase  $\gamma_{\rm ad}$ via Eq. (\ref{gamma_static}), and the value obtained dynamically using Eq. (\ref{gamma_dyn}). The blue curve shows the quantum state error, defined as $1-|\langle {\rm initial \ state} | {\rm final \ state} \rangle |$, that is, the amount of norm which becomes excited during the evolution. The state error is low ($<10^{-3}$) for 
time steps $\Delta t > 0.5$, while a similar phase error can only be achieved by significantly longer time steps $\Delta t > 15$.
}
\end{figure}

 If the duration of a time step $\Delta t$ is of the order of $1/\Delta E$, where $\Delta E$ is the energy gap, the dynamic method produces exactly the same result as the adiabatic one. Interestingly, even for much shorter time steps, the system still behaves adiabatic in the sense that its overlap with excited states remains negligible, and initial and final state remain the same up to a phase difference. However, this phase difference acquires some errors, in the sense that it differs from the adiabatic value. This behavior is demonstrated in the data shown in Fig. \ref{adi} for a system of $N=5$ electrons with Coulombic interactions and an optical potential of width $w=3 l_{\rm B}$. This finding suggests that a quasihole can be moved relatively fast without energetically exciting the system, but this yet does not guarantee an adiabatic phase evolution.
 
\section{Summary}
We have proposed several optical tools which can provide microscopic control over excitations in integer or fractional quantum Hall systems. In Sec. \ref{pump}, we have 
developed ideas for a quasiparticle pump, based on interactions between electrons and photons with orbital angular momentum. For graphene, empty and filled Landau levels can optically be coupled as discussed in Sec. \ref{graphene}. For GaAs, a spin-flip Raman coupling  is possible, see Sec. \ref{GaAs}. We have applied a STIRAP scheme on this many-body scenario, which allows avoiding decoherence.  Our techniques to create individual quasiparticles are robust against disorder and can give rise to novel ways of measuring fractional charge and statistics. A possible application within a Corbino disk geometry is given in Sec. \ref{sec:corbino}. 

In Sec. \ref{pin}, we have discussed different strategies for optically trapping a quasihole. We have studied the role played by the potential width for the stability of the trap. Even shallow potentials are found to support quasihole states, but the gap above the quasihole state is largest when the width is on the order of the magnetic length. A simple way of achieving an optical potential is based on the AC Stark shift, but the potential width, limited by the wavelength, exceeds the ideal length scale. Improvements are possible using a three-level coupling scheme, where for the prize of a weaker trap the potential width can be brought below the diffraction limit. We have also simulated the system dynamics in a moving potential, showing that such a process imprints a Berry phase in the electronic wave function according to the fractional charge of the quasihole. The optical potentials thus might become useful for braiding quasiparticles, which is the operation on which future topological quantum computers might be based on.

In summary, our manuscript advances quantum-optical tools for engineering and manipulating quantum Hall systems. In previous work, optical driving near a Landau level resonance has been suggested as a tool for engineering novel quantum Hall states \cite{areg}. Here, we have applied similar ideas in order to control bulk excitations of a quantum Hall system. Other interesting aspects of optically coupled Landau levels regard quantized dissipation rates \cite{tran18}, or optically induced electron localization \cite{arikawa17}.

\acknowledgments
We acknowledge fruitful discussions with Wade de Gottardi, Ze-Pei Cian, and Hwanmun Kim. This research was financially supported by the NSF through the PFC@JQI, AFOSR-MURI  FA95501610323, EFRI-1542863,  CNS-0958379,
CNS-0855217,  ACI-1126113  and  the  City  University  of
New  York  High  Performance  Computing  Center  at  the
College of Staten Island, Sloan Fellowship, YIP-ONR.
P.B. acknowledges support by AFOSR, NSF PFC@JQI, NSF QIS, ARL CDQI, ARO MURI, and ARO.

\appendix
\section{Ladder operators vs. projection operators in the description of a quasihole}
Whether implemented as a $\pi$-pulse or as a STIRAP process, the optical pumping described in Sec. \ref{pump} shifts an electron from orbital $m$ in the Landau level at the Fermi surface to orbital $m+1$ in an empty Landau level above the Fermi surface. In the STIRAP scheme (Sec. \ref{GaAs}), the Landau level shift is just a spin flip, so it does not modify the spatial wave function. In the $\pi$-pulse scheme (Sec. \ref{graphene}), the Landau level excitation produced by the first pulse is removed by a second pulse. So in both cases, the only effect of the pump on the spatial wave function is to modify the orbital of each electron. This effect can be described by  projection operators $\tilde b^\dagger$: $\tilde b^\dagger |m\rangle = |m+1\rangle$. As seen from Eq. (\ref{bqh}), shifting the orbital of all electrons produces a quasihole, but in  Eq.~(\ref{bqh}) these shifts are produced by the ladder operators $b^\dagger$ which also change the norm of the state, $b^\dagger |m\rangle = \sqrt{m+1} |m+1\rangle$. In this appendix we analyze the role played by these normalization factors for the many-body wave function.

For a single Slater determinant, the normalization factors are irrelevant, as they are removed by normalizing the many-body state vector. Therefore, the optical pumping is exactly the procedure which creates a hole within an integer quantum Hall state. Fractional quantum Hall states, however, consist of many Slater determinants, and each Slater determinant may obtain a different normalization factor. Then, the overall normalization of the many-body state does not fully remove the normalization factors introduced by the ladder operators. Instead, these factors will change the weight of each participating Slater determinant. 

To quantify this change, let us denote the Slater determinants by $\ket{\alpha}$. Each Slater state bijectively maps onto the Slater state $\ket{\tilde \alpha}$, in which all orbital quantum numbers are increased by one, $m_i^{(\alpha)} \leftrightarrow m_i^{(\tilde \alpha)} = m_i^{(\alpha)}+1$. We write a generic initial state as
\begin{align}
 \ket{\Psi(0)} = \sum_\alpha c_\alpha \ket{\alpha},
\end{align}
with normalized coefficients, $\sum_\alpha |c_\alpha|^2=1$. The final state then reads
\begin{align}
\label{qhtilde}
 \ket{\Psi(T)} \equiv \ket{\tilde \Psi_{\rm qh}} = \left( \prod_{i=1}^N \tilde b_i^\dagger \right) \sum_\alpha c_\alpha \ket{\alpha} =\sum_\alpha c_\alpha  \ket{\tilde\alpha},
\end{align}
In contrast, the quasihole state as defined in Eq. (\ref{bqh}) is given by
\begin{align}
\label{qhnormal}
 \ket{\Psi_{\rm qh}} &= {\cal N} \left( \prod_{i=1}^N b_i^\dagger \right) \sum_\alpha c_\alpha \ket{\alpha} \nonumber \\ 
 & = {\cal N} \sum_\alpha \sqrt{\prod_{i=1}^N (m_i^{(\alpha)}+1)}  c_\alpha \ket{\tilde \alpha}.
\end{align}
Denoting $B_\alpha \equiv  \sqrt{\prod_{i=1}^N (m_i^{(\alpha)}+1)}$, the normalization factor of the quasihole state reads ${\cal N} = \left( \sum_\alpha |c_\alpha|^2 B_\alpha^2 \right)^{-1/2}$. From this, the overlap between $\tilde \Psi_{\rm qh}$ and $\Psi_{\rm qh}$ is obtained as
\begin{align}
 |\langle \Psi_{\rm qh} | \tilde \Psi_{\rm qh} \rangle | &= \frac{ \sum_\alpha |c_\alpha|^2 B_\alpha}{ \sqrt {\sum_\alpha |c_\alpha|^2 B_\alpha^2}} = \frac{ \langle B \rangle_0}{\sqrt{\langle B^2 \rangle_0}} \nonumber \\ &
 =  \sqrt{ \frac{ \langle B \rangle_0^2}{\langle B \rangle_0^2 + {\rm var}(B)}} .
\end{align}
In the last equality, the brackets $\langle \cdot \rangle_0$ denote the quantum average of $B_\alpha$ with respect to state $\ket{\Psi(0)}$, and the variance of $B$ is taken with respect to the probability distribution described by $\ket{\Psi(0)}$. Physically relevant states are sharp in total angular momentum, and it follows that ${\rm var}(B) \ll \langle B \rangle_0^2$, so the overlap is on the order of 1. We have numerically checked this for the Laughlin state in Table \ref{Tabo}, confirming that the two different quasihole states $\ket{\Psi_{\rm qh}}$ and $\ket{\tilde \Psi_{\rm qh}}$ are approximately the same (with overlap $\sim 0.98$). Notably, their overlap grows with the system size.

As a final remark, we note that the conventional definition using the ladder operators is particularly appealing in terms of first-quantized wave functions, as the quasihole insertion simply leads to a prefactor, without modifying the structure of the polynomial wave function. On the other hand, from a second quantized point of view, the projector definition seems more natural as it leaves the coefficients $c_\alpha$ unchanged.

\begin{table}
\begin{tabular}{l|c|c|c|c|c|c|c}
 $N$ & 5 & 6 & 7 & 8 & 9 & 10 & 11 \\
 \hline
$  O $ & 0.9792 & 0.9763 & 0.9760 & 0.9769 & 0.9779 & 0.9786 & 0.9789\\
\end{tabular}
\caption{\label{Tabo} Overlap $O=|\langle \Psi_{\rm qh} |\tilde \Psi_{\rm qh} \rangle |$ between Laughlin quasihole states produced from Eq. (\ref{qhtilde}) and from Eq. (\ref{qhnormal}).}
\end{table}

%

\end{document}